\documentclass[12pt]{article}   
\usepackage{latexsym}

\normalsize
 
\if@twoside
\else
\fi
\displaywidth\textwidth %
\linewidth\textwidth %
    
\newtheorem{thm}{Theorem}[section]

\newtheorem{prop}[thm]{Proposition}

\newcommand{\nc}{\newcommand}
\nc{\sigf}{\Sigma_\phi}
\nc{\bg}{\begin{eqnarray}}
\nc{\ed}{\end{eqnarray}}
\nc{\bp}{\bar{\Psi}}
\nc{\bm}{\bar{m}}
\nc{\ol}{\overline}
\nc{\ul}{\underline}
\nc{\bee}{\begin{equation}}
\nc{\ee}{\end{equation}}
\nc{\beer}{\begin{eqnarray}}
\nc{\eer}{\end{eqnarray}}
\nc{\M}{$(M,g_{ab})\:$}
\nc{\V}{$(V^3,g_{ij})\:$}
\nc{\lap}{\bigtriangleup_2}
\nc{\dv}{\textstyle\frac}
\nc{\sdv}{\scriptstyle\frac}
\nc{\hf}{\dv{1}{2}}
\nc{\qtr}{\dv{1}{4}}
\nc{\shf}{\sdv{1}{2}}
\nc{\nn}{\nonumber}
\nc{\2}{\;\;}
\nc{\4}{\;\;\;\;}
\nc{\5}{\;\;\;\;\;}
\nc{\6}{\;\;\;\;\;\;}
\begin{document}

\vspace{0.5in}
\center{\bf \large Yang's Gravitational Theory}   
\vspace{0.5in}

\center{\bf Brendan S. Guilfoyle 
\footnote{Mathematics Department, Tralee R.T.C., Tralee, Co. Kerry, Ireland} 
and Brien C. Nolan 
\footnote{School of Mathematical Sciences, Dublin City University, Glasnevin, Dublin 9, Ireland.}}   

\vspace{0.5in}

\abstract {Yang's pure space equations (C.N. Yang, Phys. Rev. Lett. {\bf 33}, 445 (1974)) 
generalize Einstein's gravitational equations, while coming from gauge theory. We study these equations from a number of vantage points: summarizing the work done previously, comparing them with the Einstein equations and investigating their properties. In particular, the initial value problem is discussed and a number of results are presented for these equations with common energy-momentum tensors.}

\vspace{1in}




\newpage

\section{Introduction}

Over 20 years ago C. N. Yang \cite{yang} introduced a system of equations 
which, while generalizing Einstein's vacuum equations,  take the Yang-Mills
equations of gauge theory as their underlying structure.  These equations, 
which we refer to as {\it Yang's equations}, as obvious candidates 
(at the classical level at least) for the unification of gravitational and 
gauge theory, have been studied from a number of perspectives during the
subsequent years. Our intention in this paper is threefold: to summarize 
the work that has been done on these equations as they relate to classical 
general relativity, to argue that these equations are worthy of closer 
scrutiny, and, finally, to extend the work that has already been done. 
 
The basic variable in Yang's theory is, as in general relativity,  a 
Lorentz metric $g_{ab }$ on a 4-manifold $M$. The equations that this metric 
is required to satisfy are 
\bee\label{e:yang}
\nabla_a R_{bc}-\nabla_b R_{ac}=0,
\ee 
where $R_{ab}$ is the Ricci tensor and $\nabla$ is the 
covariant derivative associated with $g_{ab}$.

Yang's equations arise naturally by applying the Yang--Mills condition
$^*D^*F=0$ to the curvature 2-form (Riemann tensor) of an $so(3,1)$ valued {\em
Levi--Civita} connection. In tensorial language, this reads
$\nabla_a{R^a}_{bcd}=0$, which by the Bianchi identities, is equivalent
to (\ref{e:yang}).

We do not propose to allow these equations to supercede those of
Einstein as the field equations of space-time, but rather intend on the one
hand to investigate the consequences of Yang's equations as being of interest in their own right, and on the other to study the implications of Yang's
equation for classical general relativity.

As a third order non-linear system, Yang's equations are significantly 
more complex than those of either general relativity or Yang-Mills theory, 
both of which are second order. Nonetheless, shortly after Yang published 
his paper a number of exact solutions were discovered by Pavelle \cite{pav1} 
\cite{pav2}, Thompson \cite{thom1}\cite{thom2}, Ni \cite{ni} and Chen-Long 
{\it et al} \cite{cl1}\cite{cl2}. Later work by Aragone and Restuccia 
\cite{aar} investigated the linearized version of Yang's equations, while 
Mielke \cite{mlk1} has looked at the equations with emphasis on the double 
anti-self dual ansatz, which yields a gravitational analog of the instanton 
pseudoparticle of Yang-Mills theory.    

More recently, these equations have been subsumed into a general 
quadratic lagrangian approach, where the basic variables are the metric 
and the connection (which is not necessarily torsion free - see Baekler, 
Heyl and Mielke \cite{bhm}, Maluf \cite{mal1}\cite{mal2} and Szczyrba 
\cite{szc}). Further interest in the equations has been generated by 
suggestions that they may be suitable for describing the propagation 
of gravitational waves in a non-vacuum (see van Putten \cite{putt}).

In the next section we shall outline a number of general properties of 
solutions to Yang's equations. Many of these properties have been described 
in one form or another over the last 20 years. In addition, we study the 
relationship between the curvature orthogonality condition (a consequence 
of Yang's equations) and the algebraic structure of the Ricci tensor. 

Any classical dynamical equation is required to be deterministic, in 
the sense that, given some initial (possibly constrained) data on a 
spacelike slice of spacetime, the equations have a solution, at least 
for a short time into the future. Section 3 discusses this 
initial value problem for Yang's equations and indicates how this can be 
shown to be well-posed.

Section 4 looks at the compatability of different energy-momentum tensors 
with Yang's equations. In particular, we show that 
all Einstein--Maxwell fields obeying Yang's equations lie in Kundt's class
\cite{kundt} and all perfect fluids have Robertson--Walker geometry.

There are many open questions concerning these equations which the authors consider worth pursuing. Describe the asymptotic fall-off for YPS. Does a peeling theorem apply? Is there a hamiltonian formulation for the dynamics of the theory? Are the equations linearisation stable? How do gauge symmetries of the Levi-Civita connection relate to Killing vectors of the metric? Further study of the existing exact solutions is also warranted. 

Throughout, we will refer to Lorentz 4-manifolds \M which satisfy Yang's 
equation as {\it Yang Pure Spaces} (YPS)\footnote{The abbreviation shall be applied to both the singular and plural cases.}.

\section{Some Basic Properties of Yang Pure Spaces}

In this section, we 
review three properties of solutions of Yang's equation.
These properties have been noted previously by other authors 
(see e.g. Thompson \cite{thom1}\cite{thom2}); our aim is to present 
(and extend) these results as a preliminary step prior to obtaining 
solutions of Yang's equations  under various assumptions.

Firstly, we note that a contraction of Yang's equations (\ref{e:yang}) yields
\[ \nabla_aR^a_b-\partial_bR=0,\]
so that when this is compared with the contracted Bianchi identity
\[\nabla_aR^a_b-\hf\partial_bR=0,\]
we find that a necessary condition for \M be a YPS is that it be of constant scalar curvature.

On the other hand, because in four dimensions the full (second) 
Bianchi identities are equivalent to
\[ \nabla_a{C^a}_{bcd}=\nabla_{[c}R_{d]b}-\dv{1}{6}g_{b[d}\nabla_{c]}R,\]
Yang's equations (\ref{e:yang}) is equivalent to
\bee\label{e:1}
\nabla_a{C^a}_{bcd}=0,\qquad R = {\rm constant}.\ee
From this, we can immediately identify three large classes of solutions:
\begin{thm}
The following are sufficient (though not necessary) for \M to be a YPS:
\begin{description}
\item[(a)] \M is an Einstein space,
\item[(b)] \M is conformally flat with constant Ricci scalar,
\item[(c)] \M has parallel Ricci tensor.
\end{description}
\end{thm}
It is worth noting that Yang's equations are {\bf not} conformally 
invariant.
Indeed
\begin{thm}
Suppose that \M is a YPS.
Then a conformally related metric also solves Yang's equations 
if and only if the conformal curvature of \M is degenerate of 
type N, with twistfree repeated principal null direction.
\end{thm}
It is also worth mentioning that all solutions of class (b) 
are obtained by solving the nonlinear wave equation
\bee\label{e:2}
\Box_M\Omega+\dv{1}{6}R\Omega^3=0,\ee
for the conformal factor $\Omega$ of the metric tensor 
$g_{ab}=\Omega^2\eta_{ab}$, where 
$\Box_M$ is the d'Alembertian of flat space-time and $R$ 
is the constant Ricci scalar of $g_{ab}$. 
In fact, the existence of a solution to this equation constitutes a 
version of the ``Lorentz Yamabe problem":

\vspace{.2in}
\noindent {\bf Lorentz Yamabe Conjecture}
{\it Let \M be a closed or asymptotically flat Lorentz manifold, 
then equation (\ref{e:2}) has a solution.}
\vspace{.2in}

Note also that because of (\ref{e:1}), the Ricci and Weyl tensor terms 
in the Bianchi identities uncouple. Hence in the Newman--Penrose (NP) 
formalism, Yang's equations are equivalent to the contracted Bianchi 
identities (equations 7.69 - 7.71 of Kramer {\it et al} \cite{kram} 
or equations 4.12.40-41 of Penrose and Rindler \cite{par}), and the 
remaining Bianchi identities (equations 7.61-68 of \cite{kram} or 
equations 4.12.36-39 of \cite{par}) with either all Weyl tensor terms 
ignored and $\Lambda=R/24$ held constant, or all Ricci tensor terms ignored. 
One or other version may be used according as one has information 
regarding the Weyl or Ricci tensor.

Secondly, Yang's equations imply the equation 
\[ \nabla^{DD^\prime}\Psi_{ABCD}=0,\] for the Weyl spinor. 
Hence the classic Goldberg--Sachs theorem holds: 
\begin{thm}
In a YPS \M a null vector $k^a$ is a repeated principle null direction 
(pnd) of the Weyl tensor if and onlt if $k^a$ is tangent to a shearfree 
null geodesic. 
\end{thm}
Thus this theorem, which is an indispensible tool in the study of exact 
solutions of Einstein's equation, may also be used in the study of YPS.

Thirdly, on taking the covariant derivative of Yang's equations and 
applying the Ricci identities, one can obtain the equation
\[R_{ab[cd}{R^b}_{e]}=0,\]
which has the irreducible form
\bee\label{e:coc}
C_{ab[cd}{P^b}_{e]} =0,
\ee
where $P_{ab}\equiv R_{ab}-(R/4)g_{ab}$ is the trace-free Ricci tensor. 
Thompson [1] has pointed out some of the consequences of this equation, 
which we will refer to as the {\it Curvature Orthogonality Condition} 
(COC), for the Petrov type of non-Einstein YPS. We wish to focus on 
these consequences in more detail. 

In the section 4, we shall investigate the compatibility of 
different types of physical energy-momentum tensor with Yang's equation. 
It is therefore of interest to determine what may be deduced from 
(\ref{e:coc}), assuming a particular structure for the energy-momentum 
tensor, which determines, via Einstein's equation, the structure of the 
Ricci tensor. In particular, we will determine the Petrov types allowed 
by the COC for the different Segr\'e types of the Ricci tensor. 
In order to do so, it will be convenient  to write (\ref{e:coc}) in NP form.
There are nine independent equations as we can see from the following. 

Considering the tensor
\[ T_{abcd}\equiv C_{ae[bc}{P^e}_{d]},\]
there are four choices for each of the index sets $a$ and $bcd$. 
From the symmetries of $C_{abcd}$ and $P_{ab}$, we find that $T_{abcd}$ 
is completely trace-free yielding the six equations ${T^a}_{acd}=0$, and in 
addition obeys 
\[ T_{[abc]d}=\dv{1}{3}T_{dabc},\]
which is one further equation. There are  no further possible identities, 
so that $T_{abcd}$ has nine independent components in all. Transvecting 
with appropriate members of a null tetrad, these nine equations may be 
written 
\begin{eqnarray}
-\Phi_{21}\Psi_0+2\Phi_{22}\Psi_1+\Phi_{02}\bar{\Psi}_1-\Phi_{01}(\Psi_2+\bar{\Psi}_2) + \Phi_{00}\bar{\Psi}_3 &=&0, \label{e:a} \\
\Phi_{02}\bar{\Psi}_0 - \Phi_{20}\Psi_0 +
2\Phi_{10}\Psi_1-2\Phi_{01}\bar{\Psi}_1 - \Phi_{00}(\Psi_2-\bar{\Psi}_2) &=&
0, \label{e:b} \\
\Phi_{21}\Psi_1 -\Phi_{12}\bar{\Psi}_1 + 2\Phi_{11}(\bar{\Psi}_2 - \Psi_2)
+\Phi_{01}\Psi_3 - \Phi_{10}\bar{\Psi}_3 &=&0, \label{e:c} \\
\Phi_{22}\Psi_1 - \Phi_{12}(2\Psi_2+\bar{\Psi}_2) + \Phi_{11}\bar{\Psi}_3 +
\Phi_{02}\Psi_3 - \Phi_{10}\bar{\Psi}_4 &=&0, \label{e:d} \\
\Phi_{22}(\Psi_2-\bar{\Psi}_2) + 2\Phi_{21}\bar{\Psi}_3 - 2\Phi_{12}\Psi_3
+\Phi_{02}\Psi_4 - \Phi_{20}\bar{\Psi}_4 &=&0, \label{e:e} \\
\Phi_{22}\Psi_0 - 2\Phi_{12}\Psi_1 + \Phi_{02}(\Psi_2-\bar{\Psi}_2)
+2\Phi_{01}\bar{\Psi}_3 -\Phi_{00}\bar{\Psi}_4 &=&0. \label{e:f} 
\end{eqnarray}
Notice that equations (\ref{e:a}), (\ref{e:d}), (\ref{e:f}) are complex, 
giving two real equations each.

The standard algebraic classification of the Ricci tensor 
(Hall \cite{hall}, Kramer {\it et al} \cite{kram}, \S 5.1) 
relies upon the number and multiplicities of the eigenvalues, 
and the nature (space-like, time-like or null) and multiplicities of the
eigenvectors of the eigen-problem 
\[ {R^a}_b v^b = \lambda v^a.\]
In order to use (\ref{e:coc}) in the form (\ref{e:a}) to (\ref{e:f}), 
we will construct a null tetrad which is based on these eigenvectors and 
consider the NP components of the Ricci and Weyl tensors on this tetrad. 
A sample calculation is given below, and Tables I-IV give a complete 
description of the Petrov types allowed by the COC for a given Segr\'e type.
 In the tables, the first column indicates the Segr\'e type of the Ricci 
tensor. The second column gives the eigenvectors of the Ricci tensor, with 
time-like, null or complex eigenvectors before the comma, and eigenvectors 
with the same eigenvalue grouped in parentheses. Relations which arise for 
the NP Weyl scalars are given in the third column, and from these relations,
the allowed Petrov types are determined and indicated in the fourth column.
Where the Segr\'e type has an important physical interpretation, this is
given in the final column.  
Note that the COC provides no information about
conformally flat or Einstein spaces (Segr\'e type \{(1,111)\}), and so these
will be omitted from the discussion below. 

A Ricci tensor of Segr\'e type A1 non-degenerate (\{1,111\} in Segr\'e
notation) has a pseudo-orthonormal tetrad of eigenvectors, 
$\{ u^a,v^a_\alpha\}_{\alpha = 2,3,4}$ with corresponding eigenvalues 
$\rho_1,\rho_\alpha$, no two of which may be equal.
(The pseudo-orthonormality conditions are 
$g_{ab}v^a_\alpha v^b_\alpha=-g_{ab}u^au^b = 1$, no sum over $\alpha$.) 
From these we construct the null tetrad 
\[ k^a=\frac{(u^a+v^a_1)}{\sqrt{2}},\quad  n_a=\frac{(u^a-v^a_1)}{\sqrt{2}}, 
\quad m^a =\frac{(v^a_2+iv^a_3)}{\sqrt{2}},
\quad{\bar{m}}^a=\frac{(v^a_2-iv^a_3)}{\sqrt{2}}.\]
Then we can write 
\beer
R_{ab}&=& -(\rho_1+\rho_2)k_{(a}n_{b)} +\dv{1}{2}(\rho_2-\rho_1)
(k_ak_b+n_an_b) \nn\\
& &+ (\rho_3+\rho_4)m_{(a}{\bar m}_{b)} +
\dv{1}{2}(\rho_3-\rho_4)(m_am_b+\bm_a\bm_b).\nn
\eer
In this case, the NP components $\Phi_{00}=\Phi_{22}$ and 
$\Phi_{02}\in {\bf R}$ must be non-zero, and $\Phi_{11}$ 
may also be non-zero; all other terms are necessarily zero. 
Specializing equations (\ref{e:a}) - (\ref{e:f}) to this case, we find
immediately that $\Psi_1=\Psi_3=0$. 
Furthermore, in order to avoid degenerecy
among the eigenvalues, we must have 
$\{ \Psi_0=\Psi_4, \Psi_2\}\subset {\bf R}$. 
Then the only allowed Petrov types are I and D 
({\it cf.} \S 4.4 of Kramer {\it et al} \cite{kram}). 
The rest of the table for Class A1 Ricci tensors is deduced by 
allowing degenerecies among the eigenvalues above. 
The same form for the Ricci tensor obtains. The results are summarised in Table I.

For a Ricci tensor of class A2, there exists a complex conjugate pair of
eigenvectors $z^a_{\pm}= k^a \pm i n^a$ with eigenvalues 
$\rho_1\pm i\rho_2,\rho_2\ne 0$, and a pair of space-like 
eigenvectors $v_3, v_4$ as above.
$k^a, n^a$ are necessarily null, 
and the normalization $k_an^a=-1$ may be imposed. 
Then the Ricci tensor may be written as
\[ R_{ab} = -2\rho_1k_{(a}n_{b)} + \rho_2(k_a k_b - n_a n_b)+ 
(\rho_3+\rho_4)m_{(a}{\bar m}_{b)} +\dv{1}{2}(\rho_3-\rho_4)
(m_am_b+\bm_a\bm_b),\]
and the following Table II results. 
No physical Ricci tensor may have this structure.

For a Ricci tensor of class A3, there exists a double null eigenvector $k^a$
with eigenvalue $\rho_1$ and a pair of spacelike eigenvectors $v^a_3, v^a_4$
as above. Taking $n^a$ to complete a null tetrad in the standard way, the
Ricci tensor may be written 
\[ R_{ab} = -2\rho_1k_{(a}n_{b)} + \lambda k_ak_b +(\rho_3+\rho_4)m_{(a}{\bar
m}_{b)} + \dv{1}{2}(\rho_3-\rho_4)(m_am_b+\bm_a\bm_b),\]
where we must have $\lambda\ne 0$, for otherwise $n^a$ would be a fourth
independent eigenvector. The results are in Table III.

A class B Ricci tensor has a triple null eigenvector $k^a$ with eigenvalue
$\rho_1$ and a unique spacelike eigenvector $v^a_4$ orthogonal to $k^a$. We
complete a null tetrad with $n^a$ and $v^a_3$. Then the Ricci tensor may be
written
\[ R_{ab}=-2\rho_1k_{(a}n_{b)} + 2\sigma
k_{(a}v_{3a)}+(\rho_1+\rho_4)m_{(a}{\bar m}_{b)}
+\dv{1}{2}(\rho_1-\rho_4)(m_am_b+\bm_a\bm_b),\]
where $\sigma$ must be non-zero. Then Table IV results. There are no
physical Ricci tensors of this type.

Several results may be read off Tables I-IV. For instance we have the
following theorem.
\begin{thm}
If $k^a$ is a double (respectively triple) null eigenvector of $R_{ab}$, then
$k^a$ is at least a 2-fold (respectively 3-fold) repeated pnd of the Weyl
tensor.
\end{thm}
Note that the converse of this result is not implied by the COC. For example,
for class A2, the Weyl tensor is type D if and only if (following the results
above) $\Psi_0=\Psi_4 =0$, so that $k^a$ and $n^a$ are both 2-fold repeated
pnd's. However the Ricci tensor has no null eigenvectors.

The COC is implied by Yang's equation, but is obviously true for more general
space-times. In fact we shall see in section 4 that the condition
$\nabla_{[a}R_{b]c}=0$ imposes quite severe constraints on space-times having
certain Segr\'e types.

\section{The Initial Value Problem}

The Cauchy initial value problem for the Einstein equations of general
relativity has received much attention over the last 40 years, from the
original work of Lichnerowitz \cite{lich} and Choquet-Bruhat \cite{fb:ex}, to
the sharper results of Hughes, Kato and Marsden \cite{hkm:hyp} and the recent
work of Choquet-Bruhat and York \cite{ycby:mell}. Similarly, the initial
value problem for the Yang-Mills equations on Minkowski space has been
treated by Kerner \cite{ker:cpym}, Eardley and Moncrief \cite{eam:ge1}\cite{eam:ge2} and Klainerman and Machedon \cite{kam:fen}.Both the Einstein
and Yang-Mills equations initial value problems have been solved in somewhat
restricted situations. For the Einstein equations, short time existence only
has been established, except in certain cases. Indeed, in view of the
singularity theorems (Hawking and Ellis \cite{hae}), it appears that long
term existence is not possible in the general case. On the other hand, while long term existence has been established for the Yang-Mills equations, this has only been possible when the gauge group is compact and the base manifold is conformally related to Minkowski space-time (Choquet-Bruhat, Paneitz and Segal \cite{cb:uni}). In the case of non-compact gauge group, blow-up can occur (Yang \cite{yang:bup}) and it is not known what conditions guarantee long term existence over a general non-conformally flat Lorentz manifold. 

We do not expect Yang's equations to have a long term well-posed initial value problem for a number of reasons. In the first case, they generalize the Einstein vacuum equations, and so the singularity theorems of general relativity will also apply to them. Secondly, when viewed from a gauge-theoretic perspective, the equations lie naturally in the special orthonormal frame bundle of the base manifold, and therefore the gauge group is SO(3,1), which is non-compact. Consequently, we consider only short time existence for these equations.

We assume that the spacetime is the product of $V^3\times\Re$, where $V^3$ is a compact oriented 3-manifold and   
that the curves $\{p\}\times\Re$    
are timelike everywhere with respect to $g$, while $V_t\equiv V^3\times\{t\}$ are spacelike   
everywhere for all t.  Thus if we denote  the unit normal to $V_t$ by $\eta$   
then $g^{ab}\eta_a\eta_b=-1$ and $g$ restricted to $V_t$ is positive definite. Endow M with a smooth positive definite background  
metric $\ol{g}$ and denote the  associated Levi-Civita connection   
by $\ol{\nabla}$. Let h be $\ol{g}$ restricted to $V_0$. Define the Sobolev spaces $E_s(V^3\times I)$ to be  the space of tensors $\phi$ on   
$V^3\times$I such that   
\begin{description}   
\item[(i)]the restriction of $\phi$ and its derivatives $\ol{\nabla}^z\phi$   
of any order  $|z|\le$s to each $V_t$ is almost everywhere defined and   
square integrable in the metric h. Set    
\[\|\phi\|_{H_s(V_t)}\equiv\left(\int_{V_t}\sum_{|z|\le   
s}|\ol{\nabla}^z  \phi|^2\;d\mu(h)\right)^{\scriptstyle\frac{1}{2}}.\]   
\item[(ii)] the map I$\rightarrow \Re$ with t$\mapsto\|\phi\|_{H_{s-1}(V_t)}$ is   
continuous and  bounded, while I$\rightarrow \Re$ with   
t$\mapsto\|\phi\|_{H_s(V_t)}$ is measurable and essentially  bounded.   
\end{description}   
$E_s(V^3\times I)$ endowed with the norm   
\[\|\phi\|_{E_s(V^3\times I)}\equiv \;Ess\;sup_{t\in I}\;\|\phi\|_{H_s(V_t)},\]
is a Banach space. For $s>{\scriptstyle\frac{5}{2}}$, $E_s(V^3\times I)$ has nice embedding and multiplication properties (see Choquet-Bruhat  {\it et al} \cite{cbcf:cd}). Moreover, these are the appropriate spaces for looking at the initial value problem for quasi-linear wave equations, such as the Einstein vacuum and Yang-Mills equations. We shall use these spaces to study the initial value problem for Yang's equations.

We now find the constraints of Yang's equations on $V_0$. Denote by $^3\!g$ the pullback of $g$ to $V_0$.
The {\bf second fundamental form} K is a  symmetric 2-tensor on $V_0$ given by   
\[K_{ab}=\;^3\!g^c_a \; ^3\!g^d_b\nabla_c\eta_d.\]  
Since Yang's equations are third  order in $g_{ab}$, we also require   
certain 2nd derivatives of $g_{ab}$ to be prescribed on the initial hypersurface.
Introduce the symmetric 2-tensor $N_{ab}$ on $V_0$ which is the pullback of the   
Ricci tensor of  $g_{ab}$ to $V_0$:   
\[N_{cd}\equiv \;^3\!g^a_c \; ^3\!g_d^b R_{ab}.\]  
\begin{prop}   
The equations (\ref{e:yang}) when pulled back to $V_0$ read:    
\beer  
 D _{[c }N_{b ]a }&=&K_{a [b } D ^d K_{c ]d }-  
K_{a [b } D _{c ]}K\label{e:con1} \\    
K^d _{[a }N_{b ]d }&=& D _{[a } D ^d K_{b ]d } \label{e:con2}\\  
constant&=& ^3\!R+K^2-K_{a b }\;K^{a b }-2N^a _a  \label{e:con3}.  
\eer   
\end{prop}    
Here $K\equiv g^{ab}K_{ab}$ is the mean curvature of the embedding and $D$ is the Levi-Civita connection and $^3\!R$ the scalar curvature associated with $^3\!g$.

Now we state a theorem which establishes short time existence for Yang's equations. The details can be found in Guilfoyle \cite{guil2}.

\begin{thm} [Local Existence for the Yang's Equations]  
Let $V^3$ be a compact orientable 3-manifold and $^3\!g\in H_4(V^3)$, $K\in H_3(V^3)$, $N\in   
H_2(V^3)$ be symmetric 2- tensors on $V^3$, with $^3\!g$ positive definite. Let $^3\!g$, $K$ and $N$ satisfy the   
constraints (\ref{e:con1}), (\ref{e:con2}) and (\ref{e:con3}). Then  there   
exists $I\subset \Re$ and a unique Lorentz metric   
$\hat{g}\in E_4(V^3\times I)$ satisfying Yang's equations with   
\[\hat{g}|_{V_0}=g, \quad \partial_0 \hat{g}|_{V_0}=K \quad and \quad   
Ricci(\hat{g})|_{V_0}=N.\]  
\end{thm}

\noindent{\bf Proof}:

The proof follows that of the initial value problem for the Yang-Mills and 
Einstein equations, where the quasi-linear equations in an appropriate gauge 
are viewed as a perturbation of the corresponding linear systems. 
Then, by use of energy estimates for the linear system, a mapping 
on the spaces $E_s(V\times I)$ is shown to be a contraction for small 
enough interval $I$ (and large enough $s$). Since these are Banach spaces, 
such a map has a unique fixed point, the solution of the non-linear system.

For the Yang system of equations, a number of extra complications arise. 
Yang's equations constitute a third order quasi-linear system of partial 
differential equations for $g_{ab}$ and so fall outside of the usual (2nd 
order) wave equation framework used for the initial value problem of general 
relativity and gauge theory. In order to overcome this difficulty we can 
separate the connection from the metric and propagate them separately by 
wave equations, to which we can apply standard techniques. The key to the 
success of this method is the fact that if the connection and metric are 
initially compatible, they remain so throughout the evolution. This follows 
from a geometric identity which dictates the propogation of torsion. The 
appropriate gauge condition for these equations is a combination of the 
harmonic co-ordinate condition
\[   
g^{ce}(\hat{\Gamma}_{ce}^{d}-\ol{\Gamma}_{ce}^d)=0,
\]
and a generalized Lorentz gauge for the connection 1-forms $A$
\[   \nabla^cA_{c(a)}^{\;\;\;(b)}=0.  
\]
With this approach, the proof of the initial value problem goes through as in the Einstein case.
$\Box$

\section{The Energy-Momentum Tensor of a YPS}

As outlined above, we do not consider replacing Einstein's equation with
Yang's, but rather consider the latter to be a complementary condition which
is to be imposed on space-time. We consider YPS to be worthy of study in
their own right, but also feel it is important to investigate the interaction
of the two theories. Taking this point of view, we are led naturally to the
following question. What are the consequences of Yang's equations for the
space-times most commonly studied in general relativity?

In this section we focus on these consequences for space-times satisfying
Einstein's field equation,
\bee R_{ab}-\dv{1}{2}Rg_{ab}=8\pi T_{ab},\label{e:eeq} 
\ee
with an energy-momentum tensor having one of the following three physically
significant forms: 
\begin{description}
\item[(a)] Perfect fluid.
\bee T_{ab}=(\mu + p) u_au_b + pg_{ab},\quad u_au^a =-1. \label{e:pf} \ee
\item[(b)] Electromagnetic field.
\bee T_{ab} = F_{ac}{F_b}^c -\dv{1}{4}F_{cd}F^{cd},\quad F_{ab}=-F_{ba}.
\ee
\item[(c)] Self-interacting Scalar Field.
\bee
8\pi T_{ab}=\nabla_a\phi\nabla_b\phi-\dv{1}{2}(\nabla_c\phi\nabla^c\phi-V(\phi))g_
{ab}.
\ee
\end{description}
Combining Yang's equations with Einstein's leads to 
\bee \nabla_{[a}T_{b]c}=0.\label{e:yeq}
\ee
We consider the consequences of this equation for each of the three cases
above. The energy-momentum tensor always obeys the conservation equation
$\nabla^aT_{ab}=0$, and so (\ref{e:yeq}) implies $T=g^{ab}T_{ab}=$ constant.

\subsection{Perfect Fluid}

The condition $T=$ constant leads to the equation of state 
\bee p=\dv{1}{3}\mu + c, \label{eos} \ee
for some constant $c$, and the conservation equations for any perfect fluid
read
\[ \dot{\mu} + \theta(\mu +p) =0\]
\[ \nabla_a p +\dot{p}u_a + (\mu + p)\dot{u}_a=0,\]
where a dot indicates covariant differentiation along the fluid flow lines.
Then a straightforward calculation gives
\[ 0 = u^a \nabla_{[c}T_{b]a}
=(\mu+p)(3u_{[b}\dot{u}_{c]}-\nabla_{[c}u_{b]}).\]
Transvecting with $u^b$ then yields
\[ (\mu + p) \dot{u}_a =0,\]
and so assuming $\mu + p\neq 0$, which we shall do henceforth, we obtain
$\dot{u}_a=0$, so that the fluid flow lines are geodesic. The conservation
equations then give 
\[ \nabla_a\mu =\theta(\mu+p)u_a,\] so that the spatial gradients of both
$\mu$ and $p$ vanish.
Furthermore, this allows us to calculate
\[ 0  = \nabla_{[c}T_{b]a} = (\mu + p) (\dv{1}{3}\theta
g_{a[b}u_{c]}+u_{[b}\nabla_{c]}u_a + \nabla_{[c}u_{b]}u_a ).\]
Transvecting with $u^b$ then yields 
\bee
\nabla_cu_a=\dv{1}{3}\theta h_{ac},\label{e:vital}
\ee
where as usual $h_{ab}\equiv g_{ab}+u_au_b$ is the metric tensor on the
3-spaces orthogonal to $u^a$.  Therefore the fluid flow lines are shear-free
and twist-free as well as geodesic. It is well known (Krasi\'nski
\cite{kras}, Ellis \cite{ellis}) that these are necessary and sufficient
conditions for a perfect fluid filled space-time to have Robertson-Walker
geometry. If $\mu+p=0$, then $\mu$ and $p$ are both constant, giving $R_{ab}=
{\rm constant}\times g_{ab}$, an Einstein space.
Since (\ref{e:vital}) along with (\ref{eos}) is also a sufficient
condition on (\ref{e:pf}) to ensure equation (\ref{e:yeq}), we can
summarise as follows.
\begin{thm}\label{t:1}
A perfect fluid space-time $(M,g)$ with $\mu + p\neq 0$ is a YPS if and only
if $(M,g)$ is a Robertson-Walker space-time with $p=\dv{1}{3}\mu +c$ for
some constant $c$.
\end{thm}
This theorem may be rephrased in terms of Segr\'e types. The usual
interpretation is that a perfect fluid space-time is one for which the Ricci
tensor has Segr\'e type $\{1,(111)\}$. The energy density $\mu$ and pressure
$p$ are determined from the time-like eigenvalue $\rho_1$ and space-like
eigenvalue $\rho_2$ by $\rho_1 = -4\pi(\mu+3p)$, $\rho_2 = 4\pi(\mu-p)$. The
non-degenerecy condition $\rho_1\neq \rho_2$ is equivalent to $\mu +p\neq 0$.
In addition, $\mu$ and $p$ are usually required to satisfy cert
ain positivity requirements, the weak or dominant energy conditions (Hawking
and Ellis \cite{hae}, \S 4.3). Clearly the result above does not depend on
such conditions, and so may be restated as follows.
\begin{thm}\label{t:2}
A Segr\'e type $\{1,(111)\}$ space-time $(M,g)$ is a YPS if and only if
$(M,g)$ is a Robertson-Walker space-time with constant Ricci scalar.
\end{thm}

Comparing with the first table of \S 3 above, we see that in this case, the
COC is quite weak. In fact, since Robertson-Walker space-times are
conformally flat, there are no space-times of Petrov type $I, II, D$ or $N$
in this class.
The central relation which constrains solutions to be of Robertson-Walker
form is (\ref{e:vital}). From this equation, the Ricci identities yield
conformal flatness and 3-spaces of constant curvature. The same occurs if
there is a unit spacelike eigenvector $n^a$ of the Ricci tensor obeying 
\bee \nabla_a n_b = \dv{1}{3}(\nabla^cn_c)(g_{ab}-n_an_b). \label{e:svital}
\ee
In this case, the 3-spaces orthogonal to $n^a$ are time-like, i.e. have
Lorentzian signature. Following the steps above, it is easliy seen that
(\ref{e:svital}) holds for a Ricci tensor with Segr\'e type $\{(1,11) 1\}$
and $R=$ constant, with $n^a$ the spacelike eigenvector with distinct
eigenvalue. Thus we have the following counterpart of Theorem \ref{t:2}:
\begin{thm}\label{t:3}
A space-time $(M,g)$ with Segr\'e type $\{(1,11)1\}$ is a YPS if and only if
$(M,g)$ has constant Ricci scalar and line element 
\[ ds^2 = dz^2
+\frac{A^2(z)}{(1+\frac{k}{4}(x^2+y^2-t^2))^2} (dx^2+dy^2-dt^2), \] 
with $k$ constant.
\end{thm}
Finally, we note that the presence of a cosmological constant in  Einstein's
equation does not alter any of the results derived above. 
The only difference is that the constant $c$ in (\ref{eos}) above is in the
general case equal to $(4\lambda-R)/3$, where $\lambda$ is the cosmological
constant and $R$ is the Ricci scalar.

\subsection{Electromagnetic Fields}

We turn next to the case of a YPS generated by an electromagnetic field. The
two cases, where the electromagnetic field is non-null and where it is null
will be dealt with separately. We use the NP formalism throughout this
section.

\subsubsection{Non-null Electromagnetic Fields}

A non-null Einstein-Maxwell space-time is known as a {\it Rainich geometry} and
there are two conditions which are necessary and sufficient for space-time to
be such. These are the algebraic part,
\bee
R_{ab}{R^b}_c=\dv{1}{4}g_{ac}R_{bd}R^{bd} \neq 0, \label{e:alg}
\ee
and the analytic part,
\bee \nabla_{[a}\alpha_{b]}=0,\quad \alpha_b \equiv (R_{cd}R^{cd})^{-1}
\epsilon_{bgef}{R^g}_h\nabla^fR^{he}.
\ee
The first of these is equivalent to the Segr\'e type of the Ricci tensor
being $\{(1,1)(11)\}$, while the second follows immediately from Yang's
equation. Equation (\ref{e:alg}) allows for the existence of a cosmological constant
in (\ref{e:eeq}), so that while the trace of the energy-momentum tensor
necessarily vanishes, the Ricci scalar need not. However since we are using
the original form (\ref{e:eeq}) of Einstein's equation, we wish to rule out
this possibility, and so demand for the moment that $R=0$. We have 
then
\begin{thm}
All Segr\'e type $\{(1,1)(11)\}$ YPS with $R=0$ are non-null
Einstein-Maxwell space-times.
\end{thm}
We can explicitly determine all such space-times. In this case, there exists
a null tetrad field for which
\bee
 T_{ab} = 4\phi_1{\bar{\phi}}_1\{k_{(a}n_{b)}+m_{(a}{\bar{m}}_{b)}\}.
\label{e:nnemf} 
\ee
$k^a$ and $n^a$ are the pnd's of the electromagnetic field. Thus the only
non-vanishing Ricci tensor term is
\[ \Phi_{11}=8\pi\phi_1{\bar{\phi}}_1 > 0.\]
The field equations are the separated Bianchi identities. From these, we
immediately obtain
\[\kappa=\sigma=\lambda=\nu=\rho=\mu=\pi=\tau=0,\]
\[ \Psi_0=\Psi_1=\Psi_2= \Psi_3= \Psi_4=0,\] \[D\Phi_{11}=\Delta\Phi_{11}=\delta\Phi_{11}=0,\] so
that space-time is conformally flat (and hence by (\ref{e:1}) includes 
{\em all} conformally flat Einstein-Maxwell space-times), $\Phi_{11}$ is
constant and both pnd's of the electromagnetic field are non-diverging. Hence
the space-time lies in Kundt's class (Kramer {\it et al} \cite{kram} \S27).
There is only one such space-time, the conformally flat Bertotti-Robinson
solution (Kramer {\it et al} \cite{kram} \S 10.3), for which the line element
may be written
\bee
ds^2 = -2dudr - 2\Phi_{11}r^2du^2
+2(1+\Phi_{11}\zeta{\bar{\zeta}})^{-2}d\zeta d{\bar{\zeta}}.\label{e:lennemf}
\ee
The coordinates $x^a=\{u,r,\zeta,\bar{\zeta}\}$ are respectively a label for
the surfaces orthogonal to $k^a$ ($k_a=\nabla_a u$), an affine parameter
along the integral curves of $k^a$ 
and holomorhpic coordinates on the 2-spaces of constant curvature given by
$u,r$ constant. In fact (\ref{e:nnemf}) is the direct product of two 2-spaces
of the same constant curvature, and provides an example of a decomposable
space-time as discussed by Thompson \cite{thom2}. This space-time is static;
$\xi_a=(2\Phi_{11}r^2,1,0,0)$ is a hypersurface-orthogonal timelike Killing
vector field. 

We can also completely determine all YPS generated, via
Einstein's equation with a cosmological constant, by a non-null electromagnetic field. Here, (\ref{e:eeq}) is replaced by 
\[ R_{ab}-\dv{1}{4}Rg_{ab}+\lambda g_{ab} = 8\pi T_{ab}, \]
where $\lambda$ is the cosmological constant. Then 
\[ R= 4\lambda = {\rm constant}, \] 
and with $T_{ab}$ as in (\ref{e:nnemf}), the only difference for the spin
coefficients and curvature tensor is
\[ \Psi_2   = -\frac{R}{12} = -\frac{\lambda}{3}. \]
Again, the field equations may be completely integrated, yielding a type D
(in general) space-time which is decomposable into two 2-spaces of constant
curvature. The line element may be written as
\bee ds^2 = -2dudr - 2(\Phi_{11}-\frac{\lambda}{2})r^2du^2
+2(1+(\Phi_{11}+\frac{\lambda}{2})\zeta{\bar{\zeta}})^{-2}d\zeta
d{\bar{\zeta}}.\label{e:lenemf}
\ee
The difference between this and the conformally flat (\ref{e:lennemf}) lies
in the different values of the constant cuvature of the two 2-spaces. 

We can summarise the results for non-null electromagnetic fields as follows.
\begin{thm}
All Segr\'e type $\{(1,1)(11)\}$ YPS have line element given by
(\ref{e:lenemf}) and have $R=4\lambda$. The space-time is generated, via
Einstein's field equations with cosmological term $\lambda$, by a non-null
electromagnetic field.
\end{thm}

As with any Rainich geometry, the electromagnetic field is only determined up
to a constant duality rotation, $\phi_1\rightarrow e^{i\alpha}\phi_1$,
$\alpha$ constant. (\ref{e:lenemf}) may be generated by an electrostatic,
magnetostatic or general static electromagnetic field.

\subsubsection{Null Electromagnetic Fields and Pure Radiation}

By definition, an energy-momentum tensor which is one of these types has the
form
\[ T_{ab}=H k_ak_b, \]
for some null vector field $k^a$ and function $H$. Then on any null tetrad
based on $k^a$, we have from Einstien's equation $\Phi_{22}=8\pi H$, and all
other Ricci tensor terms vanish.
(In the presence of a cosmological term, $R=$ constant may also be non-zero.)
Equivalently, the Ricci tensor has Segr\'e type $\{(2,11)\}$ with vanishing
Ricci scalar.

From the results of the section 2, we see that the Weyl tensor must be
algebraically special with repeated pnd $k^a$, which by the Goldberg-Sachs
theorem must be shear-free and geodesic. The only other consequence of the
COC is that $\Psi_2={\bar{\Psi}}_2$ for any null tetrad preserving the
direction of $k^a$. With $\kappa=\sigma=0$, Lorentz transformations of the
null tetrad may be used to set $\epsilon = \mu-{\bar{\mu}}=0$ and
$\tau={\bar{\pi}}={\bar{\alpha}}+\beta$. Then the only field equations
(Bianchi
 identities with Weyl tensor terms ignored) which are not identically
satisfied are \[\delta\Phi_{22}=-\tau\Phi_{22},\quad
D\Phi_{22}=0={\bar{\rho}}\Phi_{22},\] and so $\rho=0$.
Thus $k^a$ is non-diverging, and the space-time lies in Kundt's class (Kramer
{\it et al} \cite{kram} \S27). For the case of a null electromagnetic field,
$\phi_2$ (the only non-zero electromagnetic field tensor term) must satisfy
the relevant Maxwell equations. These results also follow when the
cosmological constant is non-zero. Thus we have the following.
\begin{thm}
A YPS generated by a pure radiation energy-momentum tensor or a null
electromagnetic field, or equivalently one with a Ricci tensor of Segr\'e
type $\{(2,11)\}$, lies in Kundt's class. The non-diverging geodesic repeated
pnd is the double null eigenvector of the Ricci tensor, and in a null tetrad
adapted to this vector, $\Psi_2={\bar{\Psi}}_2$. 
\end{thm}

The converse of this result (``Kundt's class with null electromagnetic filed
or pure radiation energy-momentum tensor is a YPS'') is not true, as Yang's
equation gives an 'extra' Bianchi identity arising from the splitting of the
identities into homogeneous equation for the Ricci and Weyl tensors. Modulo
the implications of this extra equation, the results of \S27.6 of Kramer {\it
et al} \cite{kram} apply to this class of YPS. It may be possible to use
this extra equation to obtain further results on thes
e solutions. This is currently being investigated.

\subsection{Self-interacting Scalar Fields}

The energy momentum tensor for a self-interacting scalar field $\phi$ with
interaction potential $V(\phi)$ is given by
\bee
8\pi
T_{ab}=\phi_a\phi_b-g_{ab}(\dv{1}{2}\phi_c\phi^c-V(\phi)),\label{e:scemt}
\ee
where $\phi_a\equiv\nabla_a\phi$ and for convenience, the gravitational
constant has been absorbed into the definition of $\phi$ and $V$. The
governing equation for $\phi$ is equivalent to the vanishing divergence of
(\ref{e:scemt});
\bee
\Box\phi+V^\prime(\phi)=0. \label{e:sceq}
\ee
Then Einstein's field equation (\ref{e:eeq}) gives
\bee
R_{ab}=\phi_a\phi_b-Vg_{ab},\label{e:ricphi}
\ee 
and hence
\bee 
\phi_c\phi^c=R+4V. \label{e:prv}
\ee
Yang's equations imply that $R$ is constant. 

If $\phi_c\phi^c=0$ over some open subset of space-time, then we see from
(\ref{e:prv}) that $V=-R/4$ is constant, and consequently the Ricci tensor
has Segr\'e type $\{(2,11)\}$, as $\phi_a$ is null. This situation was dealt
with in the previous subsection, and so we assume henceforth that
$\phi_c\phi^c$ is non-zero and we can therefore restrict our attention to
open subsets of space-time on which the sign of $\phi_a\phi^a$ does not
change. We thus note immediately that Theorem \ref{t:2} and Theorem \ref{t:3} apply in this case; for $\phi_c\phi^c <0$, the Segr\'e type is
$\{1,(111)\}$, while for $\phi_c\phi^c >0$, the Segr\'e type is
$\{(1,11)1\}$. 
We now proceed to see how these geometries arise.

Following the steps of \S 4.1, we find that the covariant derivative of
$\phi_a$ is given by
\bee
\nabla_a\phi_c=3(\phi_b\phi^b)^{-1}V^\prime\phi_a\phi_c-V^\prime g_{ac}.
\label{e:second}
\ee
This equation is equivalent to (\ref{e:yeq}) for an energy-momentum tensor of
the form (\ref{e:scemt}).
The unit vector field normal to the surfaces $\sigf$ of constant $\phi$ is
\[ n^a=(\pm\phi_c\phi^c)^{-1/2}\phi^a, \]
where here and throughout, the upper sign corresponds to $\phi_c\phi^c>0$
($\sigf$ time-like), and the lower sign to $\phi_c\phi^c<0$ ($\sigf$
space-like). We find that (\ref{e:second}) leads to 
\bee
\nabla_an_b=-V^\prime(\pm\phi_c\phi^c)^{-1/2}h_{ab}, \label{e:third}
\ee
where $h_{ab}\equiv g_{ab}\mp n_an_b$ is the metric tensor of $\sigf$. 
Using (\ref{e:second}) and (\ref{e:third}), an explicit expression may be
obtained for the second covariant derivative of $n^a$. Then the Ricci
identities may be used to prove that space-time is conformally flat and that
$\sigf$ have constant curvature. In addition, we find that $V$ must obey
\bee
(R+4V)V^{\prime\prime}-3{(V^\prime)}^2=\dv{1}{3}(R+3V)(R+4V).
\label{e:const}
\ee
Equation (\ref{e:third}) allows us to introduce 'co-moving coordinates' (the
case $\phi_c\phi^c$ may be referred to as a tachyon fluid) such that
\[ n_a =\pm\nabla_a u,\quad u=x^0. \]
Surfaces of constant $u$ are surfaces of constant $\phi$, and so
$\phi=\phi(u)$, $V=V(u)$.  The possible line elements are
\bee
ds^2=\pm du^2 +A^2(u)\left(1+\dv{k}{4}(x^2+y^2\mp z^2)\right)^{-2}
\left( dx^2+dy^2\mp dz^2\right), \label{e:linels}
\ee
and $k$ may be normalized to -1, 0 or +1. The 3-spaces $\sigf$ are Lorentzian
for $\phi_c\phi^c>0$, and Riemannian for $\phi_c\phi^c<0$.

Imposing the condition $R=$ constant on (\ref{e:linels}) yields a
differential equation for $A(u)$ which may be integrated to give
\begin{eqnarray}\label{e:asol} 
A^2(u)&=&-6kR^{-1}+a_0\exp\left(\sqrt{\pm\dv{R}{3}}u\right)
+a_1\exp\left(-\sqrt{\pm\dv{R}{3}}u\right), \quad (R\neq 0)\label{e:asola}\\
A^2(u)&=&\pm ku^2+a_0u+a_1,\quad (R=0). \label{e:asolb}
\end{eqnarray}
In (\ref{e:asolb}) a coordinate transformation of the form 
$(u\rightarrow u +$ constant has been used to set a constant of integration to zero.
(\ref{e:linels}) along with (\ref{e:asol}) with the lower sign gives the
possbible line elements for a YPS with a perfect fluid energy-momentum tensor.

The scalar field $\phi$ and potential $V$ may be determined as follows. 
Using a dot to indicate differentiation with respect to $u$, we have 
\[ \phi_a = \pm\dot{\phi} n_a,\]
and so using (\ref{e:prv}),
\[V^\prime(\phi)=\pm\dv{1}{2}\ddot{\phi}.\]
In the coordinates of (\ref{e:linels}), we find
\[ \Box\phi=\pm(\ddot{\phi}+3A^{-1}\dot{A}\dot{\phi}), \]
and so (\ref{e:sceq}) gives
\[ \ddot{\phi}+2A^{-1}\dot{A}\dot{\phi}=0,\]
which has the first integral
\bee
\dot{\phi}=aA^{-2}(u),\label{e:pdot} 
\ee
where $a$ is a constant. Thus by (\ref{e:prv}),
\[ V=\pm\dv{a^2}{4}A^{-4}-\dv{R}{4}.\]
Then evaluating the terms in (\ref{e:const}), we find that $A$ must satisfy
\bee \ddot{A}=-\dv{1}{4}a^2A^{-3}\pm\dv{R}{12}A. \label{e:aeq2}
\ee
We compare this equation with (\ref{e:asol}) to determine $a$.

For $R=0$, we find that $a=a_0$. (\ref{e:pdot}) may be integrated and we find
\[ \phi =\ln{\left| \frac{u}{\pm ku +a_0} \right| },
\qquad V=\pm\frac{a_0^2}{4}(\pm ku^2 +a_0 u)^{-2}. \]

For $R\neq 0$, comparison of (\ref{e:aeq2}) and (\ref{e:asola}) yields 
\[ a^2 =\pm 12R^{-1}\mp \dv{4}{3}a_0a_1 R.\]
$\phi$ is determined via (\ref{e:pdot}) and (\ref{e:asola}) by an elliptic
integral.

Summarising the main results, we have the following.
\begin{thm}
A YPS generated by a self-interacting scalar field $\phi$ obeying $\nabla_a
\phi\nabla^a\phi\neq 0$ has line element
\[ ds^2 =\epsilon du^2 +A^2(u)(1+\dv{k}{4}(x^2+y^2-\epsilon
z^2))^{-2}(dx^2+dy^2-\epsilon dz^2), \] 
where $\epsilon = {\rm sgn}(\nabla_a \phi\nabla^a\phi)$ and $A^2(u)$ is given
by (\ref{e:asol}).
\end{thm}

\section{Conclusions}

Perhaps the most important result reported here is that the short-time
initial value probelm for Yang's equations is well-posed. 
This is a prerequisite for YPS to be worthy of study as physical models,
and has not been demonstrated before \cite{guil2}. Other authors have
emphasised their criticisms of Yang's equations as the fundamental
gravitational field equations \cite{pav2}, \cite{pav3}, \cite{thom2}. This
has been done principally on the basis that Birkhoff's theorem does not apply
to YPS; source-free ($\nabla_{[a}R_{b]c}=0$) spherically symmetric
solutions are not necessarily static, and indeed the static spherical
source-free solutions form a four parameter family, rather than the one parameter
(Schwarzschild mass) family of general relativity. Consequently, one can
easily produce `vacuum' spherically symmetric solutions, which are candidates
for the gravitational field of the sun, but which display unphysical
characteristics - lack of gravitational red-shift, lack of bending of
starlight, and incorrect values (and direction!) for the perihelion shift. 

We have attempted here to establish an alternative philosophy for Yang's 
equations: they
are to be used in conjunction with Einstein's equation and/or appropriate
boundary conditions in our description of space-time. In this way, well established results of relativistic
astrophysics and cosmology may be maintained. 
Our hope is that a study of these equations may shed further light on the
classical gauge theoretic structure of gravity and in turn classical gauge
theory in general. For example the link between a symmetry of a YPS as a
gauge configuration and as a space-time may yield insight into the question
of how the former is to be defined \cite{traut}. This work is to be
undertaken in the future; we are currently investigating the structure of
static YPS. Details on the short-time initial value problem will appear
 elsewhere.

\begin{thebibliography}{99}
\bibitem{yang}
C.N. Yang (1974).
Phys. Rev. Lett., {\bf 33}, 445.

\bibitem{pav1}
R. Pavelle (1974).
Phys. Rev. Lett., {\bf 33}, 1461.

\bibitem{pav2}
R. Pavelle (1975).
Phys. Rev. Lett., {\bf 34}, 1114.

\bibitem{thom1}
A.H. Thompson (1975).
Phys. Rev. Lett., {\bf 35}, 320.

\bibitem{thom2}
A.H. Thompson (1975).
Phys. Rev. Lett., {\bf 34}, 507.

\bibitem{ni}
W.-T. Ni (1975).
Phys. Rev. Lett., {\bf 35}, 319.

\bibitem{cl1}
C. Chen-lung, G. Han-ying, C. Shi and H. Zuo-xiu (1976).
Acta Astr. Sinica, {\bf 17}, 147.

\bibitem{cl2}
C. Chen-lung, G. Han-ying, C. Shi and H. Zuo-xiu (1977).
Chinese Astronomy, {\bf 1}, 292.

\bibitem{aar}
C. Aragone and A. Restuccia (1978).
Gen. Rel. Grav., {\bf 9}, 409.

\bibitem{mlk1}
E.M. Mielke (1981).
Gen. Rel. Grav., {\bf 13},  175.

\bibitem{bhm}
P. Baekler, F. W. Hehl and E. W. Mielke (1982).
In {\em Proceedings of the 2nd {M}arcel {G}rossmann meeting on general relativity,} {North-Holland, Amsterdam.} p. 413.

\bibitem{mal1}
J. Maluf (1988).
Classical Quantum Gravity, {\bf 5}, L81.

\bibitem{mal2}
J. Maluf (1991).
J. Math. Phys., {\bf 32}, 1556.

\bibitem{szc}
V. Szczyrba (1987).
Phys. Rev. D, {\bf 3}, 351.

\bibitem{putt}
M. H. P. M. van Putten (1994).
In {\em Proceedings of the November 6-8 Meeting of the Grand
                  Challenge Alliance on Black Hole Collisions},
(ed. E. Seidel), {NCSA}.

\bibitem{kundt}
W. Kundt (1961).
Z. Phys.,{\bf 163}, 77.

\bibitem{kram}
D. Kramer, H. Stephani, E. Herlt, M. MacCallum and E. Schmutzer (1980).
{\em Exact Solutions of Einstein's Field Equations.}
Cambridge University Press, Cambridge.

\bibitem{par}
R. Penrose and W. Rindler (1982).
{\em Spinors and Spacetime.}
Cambridge University Press, Cambridge.

\bibitem{hall}
G. S. Hall (1982).
Banach Center Publications, Volume 12.
Polish Scientific Publishers, Warsaw.

\bibitem{lich}
A. Lichnerowicz (1967).
{\em Relativistic Hydrodynamics and Magnetohydrodynamics.}
Benjamin, New York.

\bibitem{fb:ex}
Y. Foures-Bruhat (1952).
Acta Math., {\bf  88}, 141.

\bibitem{hkm:hyp}
T. Hughes, T. Kato and J. Marsden (1976).
Arch. Rat. Mech. and Anal., {\bf 63}, 276.

\bibitem{ycby:mell}
Y. Choquet-Bruhat and J. York (1997).
In {\em Gravitation, Electromagnetism and Geometric Structures}(ed. G. Ferrarese), Pythagora Editrice. 

\bibitem{ker:cpym}
R. Kerner (1974)
Ann. Inst. Henri Poincar\,e, {\bf XX 3}, 279.

\bibitem{eam:ge1}
D. M. Eardley and V. Moncrief (1982).
Commun. Math. Phys., {\bf 83}, 171.

\bibitem{eam:ge2}
D. M. Eardley and V. Moncrief (1982).
Commun. Math. Phys., {\bf 84}, 193.

\bibitem{kam:fen}
S. Klainerman and M. Machedon (1995).
Annals of Math., {\bf 142}, 39.

\bibitem{hae}
S.W. Hawking, G.F.R. Ellis (1973).
{\em The Large Scale Structure of Spacetime.}
Cambridge University Press, Cambridge.

\bibitem{cb:uni}
Y. Choquet-Bruhat and S. M. Paneitz and I. E. Segal (1983)
J. Funct. Anal., {\bf 53}, 112.

\bibitem{yang:bup}
Y. Yang (1990).
J. Math. Phys., {\bf 31}, 1237.

\bibitem{cbcf:cd}
Y. Choquet-Bruhat, D. Christodoulou and M. Francaviglia (1978).
Ann. Inst. Henri Poincar\,e,
{\bf XXIX 3}, 241.

\bibitem{guil2}
B. Guilfoyle (1997).
{\em The {C}auchy initial value problem for {Y}ang-{M}ills metrics.}
PhD thesis, University of Texas at Austin, Austin, Texas.

\bibitem{kras}
A. Krasi\'nski (1997).
{\em Inhomogenous cosmological models.}
Cambridge University Press, Cambridge.

\bibitem{ellis}
G. F. R. Ellis (1971).
In
{\em Proceedings of the International School of Physics "{E}nrico {F}ermi", General Relativity and Cosmology} (ed. R. K. Sachs),
Academic Press, New York.

\bibitem{pav3}
R. Pavelle  (1976).
Phys. Rev. Lett., {\bf 37}, 961.


\bibitem{traut}
A. Trautman (1979).
Bull. Acad. Polon. Sci. Series 9 Sci. Phys. Astronom.,
{\bf 27}, 7.

\end{thebibliography}

\newpage
\pagestyle{empty}
\vspace{0.5in}
\center{\bf \large Yang's Gravitational Theory}   
\vspace{0.5in}
\addtocounter{footnote}{-3}
\center{\bf Brendan S. Guilfoyle 
\footnote{Mathematics Department, Tralee R.T.C., Tralee, Co. Kerry, Ireland} 
and Brien C. Nolan 
\footnote{School of Mathematical Sciences, Dublin City University, Glasnevin, Dublin 9, Ireland.}}   

\vspace{0.5in}


\vspace{1in}

\noindent{\bf Running head:} Yang's gravitational theory.

\vspace{0.2in}
\noindent{\bf Keywords:} Initial value problem, energy-momentum tensor.

\vspace{0.2in}

\noindent{\bf Review proofs to:} 
Dr. Brendan Guilfoyle, Tralee R.T.C., Tralee, Co. Kerry, Ireland.
Phone: +353-66-24666 e-mail:bguilfoyle@staffmail.rtc-tralee.ie
\newpage
\begin{center}
{\bf Table I: Class A1 Ricci Tensor}
\vskip .2truein
\begin{tabular}{|c|c|c|c|c|} \hline
Segr\'e Notation & Eigenvectors & Weyl Scalars & Allowed & 
Physical \\ 
& & & Petrov Types & Interpretation \\ \hline
\{1,111\} & \{$u,v_2\,v_3\,v_4$\} & $\Psi_1=\Psi_3=0$, & I or D &  \\
                &                              & $\{\Psi_0=\Psi_4, \Psi_2
\}\subset ${\bf R} &      &  \\ \hline
\{(1,1)11\} & \{$(u,v_2)\,v_3\,v_4$\} & $\Psi_1=\Psi_3=0$, & I II D or N & 
\\
                &                              & $\{\Psi_0,\Psi_4,
\Psi_2\}\subset ${\bf R} &      &  \\ \hline
\{1,1(11)\} & \{$u,v_2\,(v_3\,v_4)$\} & $\Psi_1=\Psi_3=0$, & I or D &  \\
                &                              &${\bar\Psi}_0=\Psi_4, \Psi_2
\in ${\bf R} &      &  \\ \hline
\{(1,1)(11)\} & \{$(u,v_2) (v_3\,v_4)$\} & $\Psi_1=\Psi_3=0$, & I II D or N &
Non-null  \\
                &                              & $\Psi_2 \in ${\bf R} &     
& em field.\\ \hline
\{1,(111)\} & \{$u,(v_2\,v_3\,v_4)$\} &
$\Psi_3=-{\bar\Psi}_1,\Psi_4={\bar\Psi}_0$& I II D or N & Perfect fluid; \\
                &                              &  $\Psi_2 \in$ {\bf R} &     
& scalar field. \\ \hline
\{ (1,11)1\} & \{ $(u,v_2\,v_3)\,v_4 $\} & $\Psi_A \in$ {\bf R}, A =0 - 4. &
I II D III or N & Scalar field. \\ \hline
\end{tabular}
\end{center} 
\newpage
\begin{center}
{\bf Table II: CLASS A2 Ricci Tensor}
\vskip .2truein
\begin{tabular}{|c|c|c|c|} \hline
Segr\'e Notation & Eigenvectors & Weyl Scalars & Allowed Petrov Types \\
\hline
\{$ z {\bar z}, 11$ \} & \{$ z_+ z_- , v_3 v_4 $\} & $\Psi_1=\Psi_3=0 $& I or
D \\
                        &                            & $\{ \Psi_4 =
-\Psi_0,\Psi_2\}\subset ${\bf R}& \\ \hline \{ $z {\bar z}, (11) $\} & \{
$z_+ z_- , (v_3 v_4) $\} & $\Psi_1=\Psi_3=0$ & I or D \\                     
  &                            &  $\{\Psi_4 =-{\bar\Psi}_0, \Psi_2\}\in ${\bf
R}& \\ \hline
\end{tabular}
\end{center}
\newpage
\begin{center}
{\bf Table III CLASS A3 Ricci Tensor}
\vskip .2truein
\begin{tabular}{|c|c|c|c|c|} \hline
Segr\'e Notation & Eigenvectors & Weyl Scalars & Allowed  & Physical  \\ 
                 &              &              & Petrov Types         &
Interpretation. \\ \hline             
\{2,11\} & \{$k, v_3\, v_4$\} & $\Psi_0=\Psi_1=\Psi_3=0$, & II D or N & \\   
      &                & $\{\Psi_2,\Psi_4\}\subset ${\bf R} &  & \\   \hline
\{2,(11)\} & \{$k, (v_3\, v_4)$\} & $\Psi_0=\Psi_1=\Psi_3=0$, & II D or N & \\
           &                  & $\Psi_2 \in ${\bf R}      &           & \\
\hline
\{(2,1)1\} & \{$(k, v_3)\, v_4$\} & $\Psi_0=\Psi_1=0$, & II D III or N & \\ 
         &                & $\{\Psi_2,\Psi_3, \Psi_4\}\subset ${\bf R} &  &
\\  \hline
\{(2,11)\} & \{$(k,v_3\, v_4)$\} & $\Psi_0=\Psi_1=0 $& II D III or N &Null em
field, \\
           &                 &  $\Psi_2 \in$ {\bf R} &           &pure
radiation. \\ \hline
\end{tabular}
\end{center}
\newpage
\begin{center}
{\bf Table IV: CLASS B Ricci Tensor}
\vskip .2truein
\begin{tabular}{|c|c|c|c|}\hline
Segr\'e Notation & Eigenvectors & Weyl Scalars & Allowed Petrov Types \\
\hline 
\{3,1\} & \{$k, v_4$\} & $\Psi_0=\Psi_1=\Psi_2=0$, & III or N \\        &    
       & $\{\Psi_3,\Psi_4\}\subset ${\bf R} & \\ \hline \{(3,1)\} & \{$(k,
v_4)$\} & $\Psi_0=\Psi_1=\Psi_2=0$, & III or N \\        &            &
$\Psi_3\in ${\bf R} & \\ \hline
\end{tabular}
\end{center}
\end{document}